\documentclass[aps,prb,twocolumn,showpacs,superscriptaddress,floatfix,10pt]{revtex4-1}

\usepackage{graphicx}
\usepackage{amsmath}
\usepackage{epsfig}
\usepackage{helvet}
\usepackage{amssymb}
\usepackage{url}

\newcommand{\CC}{C\nolinebreak\hspace{-.05em}\raisebox{.4ex}{\tiny\bf +}\nolinebreak\hspace{-.10em}\raisebox{.4ex}{\tiny\bf +}}
\def\CC{{C\nolinebreak[4]\hspace{-.05em}\raisebox{.4ex}{\tiny\bf ++}}}

\begin{document}

\title{A Practical Guide to the Numerical Implementation of Tensor Networks I: \\ Contractions, Decompositions and Gauge Freedom}

\author{Glen Evenbly}
\affiliation{School of Physics, Georgia Institute of Technology, Atlanta, GA 30332, USA}
\email{glen.evenbly@gmail.com}
\date{\today}

\begin{abstract}
We present an overview of the key ideas and skills necessary to begin implementing tensor network methods numerically, which is intended to facilitate the practical application of tensor network methods for researchers that are already versed with their theoretical foundations. These skills include an introduction to the contraction of tensor networks, to optimal tensor decompositions, and to the manipulation of gauge degrees of freedom in tensor networks. The topics presented are of key importance to many common tensor network algorithms such as DMRG, TEBD, TRG, PEPS and MERA. 
\end{abstract}


\maketitle


\section{Introduction} \label{sect:Intro}
Tensor networks have been developed as a useful formalism for the theoretical understanding of quantum many-body wavefunctions \cite{TN1, TN2, TN3, TN4, TN5, TN6, TN7, TN8, TN9, TN10}, especially in regards to entanglement \cite{Ent1, Ent2, Ent3}, and are also applied as powerful numeric tools and simulation algorithms. Although developed primarily for the description of quantum many-body systems, they have since found use in a plethora of other applications such as quantum chemistry \cite{Chem1, Chem2, Chem3, Chem4, Chem5}, holography \cite{Holo1, Holo2, Holo3, Holo4, Holo5, Holo6}, machine learning \cite{ML1, ML2, ML3, ML4, ML5} and the simulation of quantum circuits \cite{QC1, QC2, QC3, QC4, QC5, QC6}.

There currently exist many useful references designed to introduce newcomers to the underlying theory of tensor networks \cite{TN1, TN2, TN3, TN4, TN5, TN6, TN7, TN8, TN9, TN10}. Similarly, for established tensor network methods, there often exist instructional or review articles that address the particular method in great detail \cite{Rev1, Rev2, Rev3, Rev4, Rev5}. Nowadays, many research groups have also made available tensor network code libraries \cite{TB1, TB2, TB3, TB4, TB5, TB6, TB7, TB8}. These libraries typically allow other researchers to make use of highly optimised tensor network routines for practical purposes (such as for the numerical simulation of quantum many-body systems). 

Comparatively few are resources intended to help researchers that already possess a firm theoretical grounding to begin writing their own numerical implementations of tensor network codes. Yet such numerical skills are essential in many areas of tensor network research: new algorithmic proposals typically require experimentation, testing and bench-marking using numerics. Furthermore, even researchers solely interested in the application of tensor network methods to a problem of interest may be required to program their own version of a method, as a pre-built package may not contain the necessary features as to be suitable for the unique problem under consideration. The purpose of our present work is to help fill this aforementioned gap: to aid students and researchers, who are assumed to possess some prior theoretical understanding of tensor networks, to learn the practical skills required to program their own tensor networks codes and libraries. Indeed, our intent is to arm the interested reader with the key knowledge that would allow them to implement their own versions of algorithms such as the density matrix renormalization group (DMRG) \cite{DMRG1, DMRG2, DMRG3}, time-evolving block decimation (TEBD) \cite{TEBD1,TEBD2}, projected entangled pair states (PEPS) \cite{PEPS1, PEPS2, PEPS3}, multi-scale entanglement renormalization ansatz (MERA) \cite{MERA1}, tensor renormalization group (TRG) \cite{TRG1, TRG2} or tensor network renormalization (TNR) \cite{TNR1}. Furthermore, this manuscript is designed to compliment online tensor network tutorials \cite{TenNet}, which have a focus on code implementation, with more detailed explanations on tensor network theory. 
\section{Preliminaries} \label{sect:Prelim}
\subsection{Prior knowledge}
\begin{figure}[!t!b]
\begin{center}
\includegraphics[width=8cm]{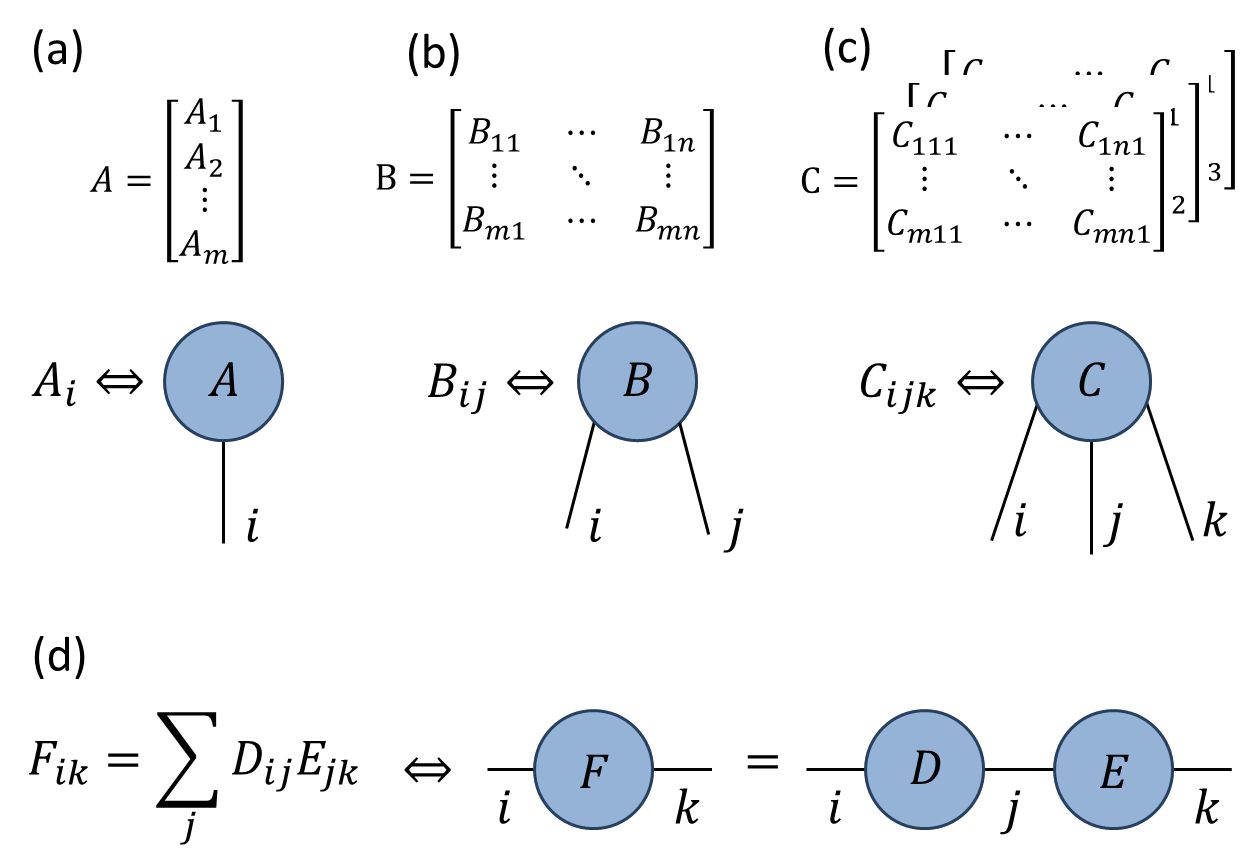}
\caption{(a-c) Diagrammatic representations of a vector $A_i$ (or order-$1$ tensor), a matrix $B_{ij}$ (or order-$2$ tensor) and an order-$3$ tensor $C_{ijk}$. (d) A contraction, or summation over an index, between two tensors is represented by a line joining two tensors.} \label{fig:S0}
\end{center}
\end{figure}

As stated above, the goal of this manuscript is to help readers that already possess some understanding of tensor network theory to apply this knowledge towards numeric calculations. Thus we assume that the reader has some basic knowledge of tensor networks, specifically that they understand what a tensor network is and have some familiarity with the standard diagrammatic notation used to represent them. An overview of these concepts is presented in Fig. \ref{fig:S0}, otherwise more comprehensive introductions to tensor network theory can be found in Refs. [\onlinecite{TN1, TN2, TN3, TN4, TN5, TN6, TN7, TN8, TN9, TN10}].

Note that we shall not assume prior knowledge of quantum many-body physics, which is the most common application of tensor network algorithms. The skills and ideas that we introduce in this manuscript are intended to be general for the tensor network formalism, rather than for their use in a specific application, thus can also carry over to other area in which tensor networks have proven useful such as quantum chemistry \cite{Chem1, Chem2, Chem3, Chem4, Chem5}, holography \cite{Holo1, Holo2, Holo3, Holo4, Holo5, Holo6}, machine learning \cite{ML1, ML2, ML3, ML4, ML5} and the simulation of quantum circuits \cite{QC1, QC2, QC3, QC4, QC5, QC6}.

\subsection{Software Libraries}
Currently there exists a wide variety of tensor network code libraries, which include Refs. \onlinecite{TB1, TB2, TB3, TB4, TB5, TB6, TB7, TB8}. Many of these libraries differ greatly in not only their functioning but also their intended applications, and may have their own specific strengths and weaknesses (which we will not attempt to survey in the present manuscript). Almost all of these libraries contain tools to assist in the tasks described in this manuscript, such as contracting, decomposing and re-gauging tensor networks. Additionally many of these libraries also contain full featured versions of complete tensor network algorithms, such as DMRG or TEBD. For a serious numerical calculation involving tensor networks, one where high performance is required, most researchers would be well advised to utilise an existing library. 

However, even if ultimate intent is to use existing library, it is still desirable that one should understand the fundamental tensor network manipulations used in numerical calculations. Indeed, this understanding is necessary to properly discern the limitations of various tensor network tools, to ensure that they are applied in an appropriate way, and to customize the existing tools if necessary. Moreover, exploratory research into new tensor network ansatz, algorithms and applications often requires non-standard operations and tensor manipulations which may not be present in any existing library, thus may require the development of extensive new tensor code. In this setting it can be advantageous to minimize or to forgo the usage of an existing library (unless one was already intimately familiar with its inner workings), given the inherent challenge of extending a library beyond its intended function and the possibility of unintended behavior that this entails.

In the remaining manuscript we aim to describe key tensor network operations (namely contracting, decomposing and re-gauging tensor networks) with sufficient detail that would allow the interested reader to implement tasks numerically without the need to rely on an existing code library.

\subsection{Programming language}
Before attempting to implement tensor methods numerically one must, of course, decide on which programming language to use. High-level languages with a focus on scientific computation, such as MATLAB, Julia and Python (with Numpy) are well-suited for implementing tensor network methods as they have native support for multi-dimensional arrays (i.e. tensors), providing simple and convenient syntax for common operations on these arrays (indexing, slicing, scalar operations) as well as providing a plethora of useful functions for manipulating these tensor objects. Alternatively, some tensor network practitioners may prefer to use lower-level languages such as Fortran or \CC{} when implementing tensor network algorithms usually for the reason of maximising performance. However, in many tensor network codes the bulk of the computation time is spent performing large matrix-matrix multiplications, for which even interpreted languages (like MATLAB) still have competitive performance with compiled languages as they call the same underlying BLAS routines. Nonetheless, there are some particular scenarios, such as in dealing with tensor networks in the presence of global symmetries \cite{Sym1, Sym2, Sym3, Sym4, Sym5, Sym6, Sym7}, where a complied language may offer a significant performance advantage. In this circumstance Julia, which is a compiled language, or Python, in conjunction with various frameworks which allow it to achieve some degree of compilation, may be appealing options.



\subsection{Terminology}
Before proceeding, let us establish the terminology that we will use when discussing tensor networks. We define the \emph{order} of a tensor as the number of indices it has, i.e. such that a vector is order-$1$ and a matrix is order-$2$. The term \emph{rank} (or decomposition rank) of a tensor will refer to the number of non-zero singular values with respect to a some bi-partition of the tensor indices. Note that many researchers also use the term \emph{rank} to describe the number of tensor indices; here we use the alternative term \emph{order} specifically to avoid the confusion that arises from the double usage of \emph{rank}. The number of values a tensor index can take will be referred to as the dimension of an index (or bond dimension), which is most often denoted by $\chi$ but can also be denoted by $m$, $d$ or $D$. In most cases, the use of $d$ or $D$ to denote a bond dimension is less preferred, as this can be confused which the spatial dimension of the problem under consideration (e.g. when considering a model on a $1D$ or $2D$ lattice geometry). 
 

\begin{figure}[!t!b]
\begin{center}
\includegraphics[width=7cm]{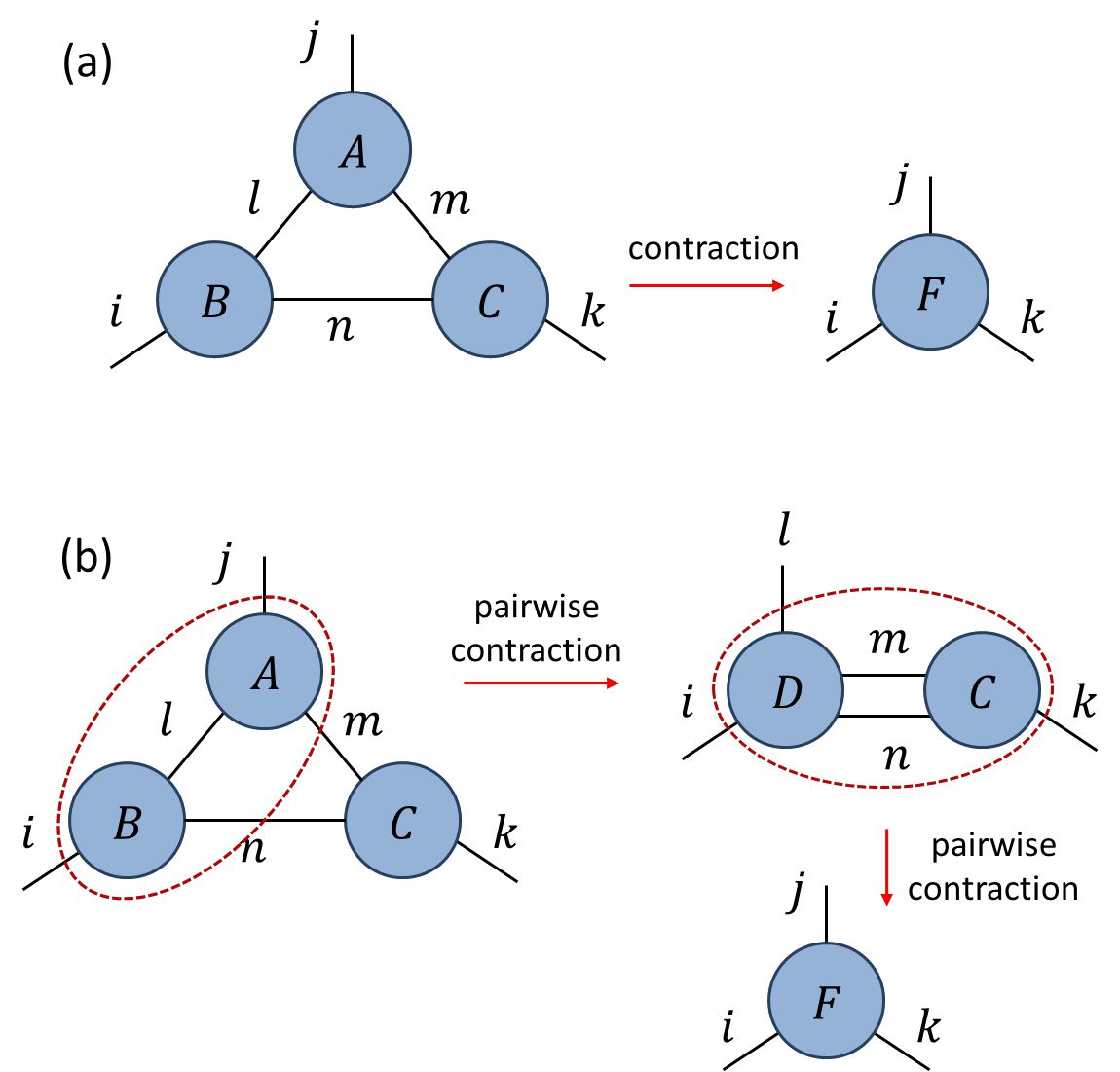}
\caption{(a) The internal indices $(l, m, n)$ of the network $\{A,B,C \}$ are contracted to give tensor $F$. (b) The network is contracted via a sequence of two pairwise tensor contractions, the first of which results in the intermediate tensor $D$.}
\label{fig:S1}
\end{center}
\end{figure}

\section{Tensor Contractions} \label{sect:Contract}
The foundation of all tensor networks routines is the contraction of a network containing multiple tensors into a single tensor. An example of the type problem that we consider is depicted in Fig. \ref{fig:S1}(a), where we wish to contract the network of tensors $\{A,B,C\}$ to form an order-$3$ tensor $F$, which has components defined 
\begin{equation}
{F_{ijk}} = \sum\limits_{l,m,n} {{A_{ljm}}{B_{iln}}{C_{nmk}}}. \label{eq:s1}
\end{equation}
Note that a convention for tensor index ordering is required for the figure to be unambiguously translated to an equation; here we assumed that indices progress clock-wise on each tensor starting from 6 o'clock. Perhaps the most obvious way to evaluate Eq. \ref{eq:s1} numerically would be through a direct summation over the indices $(l, m, n)$, which could be implemented using a set of nested `\texttt{FOR}' loops. While this approach of summing over all internal indices of a network will produce the correct answer, there are numerous reasons why this is not the preferred approach for evaluating tensor networks. The foremost reason is that it is not the most computationally efficient approach (excluding, perhaps, certain contractions involving sparse tensors, which we will not consider here). Let us analyse the contraction cost for the example given in Eq. \ref{eq:s1}, assuming all tensor indices are $\chi$-dimensional. A single element of tensor $F$, which has $\chi^3$ elements in total, is given through a sum over indices $(l, m, n)$, which requires $O(\chi^3)$ operations. Thus the total cost of evaluating tensor $F$ through a direct summation over all internal indices is $O(\chi^6)$.

Now, let us instead consider the evaluation of tensor $F$ broken up into two steps, where we first compute an intermediate tensor $D$ as depicted in Fig.\ref{fig:S1}(b),
\begin{equation}
{D_{ijmn}} = \sum\limits_l {{A_{ljm}}{B_{iln}}}, \label{eq:s2} 
\end{equation}
before performing a second contraction to give the final tensor $F$,
\begin{equation}
{F_{ijk}} = \sum\limits_{n,m}^{} {{D_{ijmn}}} {C_{nm}}. \label{eq:s3}
\end{equation}
Through similar logic as before, one finds that the cost of evaluating intermediate tensor $D$ scales as $O(\chi^5)$, whilst the subsequent evaluation of $F$ in Eq. \ref{eq:s3} is also $O(\chi^5)$. Thus breaking the network contraction down into a sequence of smaller contractions each only involving a pair of tensors (which we refer to as a \emph{pairwise tensor contraction}) is as computationally cheap or cheaper for any non-trivial bond dimension $(\chi>1)$. This is true in general: for any network of 3 or more (dense) tensors it is always at least as efficient (and usually vastly more efficient) to break network contraction into sequence of pairwise contractions, as opposed to directly summing over all the internal indices of the network. 

Two natural questions arise at this point. (i) What is optimal way to implement a single pairwise tensor contraction? (ii) Does the chosen sequence of pairwise contractions affect the total computational cost and, if so, how does one decide what sequence to use? We begin by addressing the first question.

\subsection{Pairwise tensor contractions}
Let us consider the problem of evaluating a pairwise tensor contraction, denoted $(A \times B)$, between tensors $A$ and $B$ that are connected by one or more common indices. A straight-forward method to evaluate such contractions, as in the examples of Eq. \ref{eq:s2} and Eq. \ref{eq:s3}, is by using nested `\texttt{FOR}' loops to sum over the shared indices. The computational cost of this evaluation, in terms of the number of scalar multiplications required, can be expressed concisely as   
\begin{equation}
\textrm{cost}:(A \times B) = \frac{{\left| {\dim (A)} \right| \cdot \left| {\dim (B)} \right|}}{{\left| {\dim (A \cap B)} \right|}}, \label{eq:s4}
\end{equation}
with $\left| {\dim (A)} \right|$ as the total dimension of $A$ (i.e. the product of its index dimensions) and $\left| {\dim (A \cap B)} \right|$ as the total dimension of the shared indices.

Alternatively, one can recast a pairwise contraction as a matrix multiplication as follows: first reorder the free indices and contracted indices on each of $A$ and $B$ such that they appear sequentially (which can be achieved in MATLAB using the `\texttt{permute}' function) and then group the free-indices and the contracted indices each into a single index (which can be achieved in MATLAB using the `\texttt{reshape}' function). After these steps the contraction is evaluated using a single matrix-matrix multiplication, although the final product may also need to be reshaped back into a tensor. Recasting as a matrix multiplication does not reduce the formal computational cost from Eq. \ref{eq:s4}. However, modern computers, leveraging highly optimized BLAS routines, typically perform matrix multiplications significantly more efficiently than the equivalent `\texttt{FOR}' loop summations. Thus, especially in the limit of tensors with large total dimension, recasting as a matrix multiplication is most often the preferred approach to evaluate pairwise tensor contractions, even though this requires some additional computational overhead from the necessity of rearranging tensor elements in memory when using `\texttt{permute}'. Note that the `\texttt{tensordot}' function in the Numpy module for Python conveniently evaluates a pairwise tensor contraction using this matrix multiplication approach.

\begin{figure}[!t!b]
\begin{center}
\includegraphics[width=8cm]{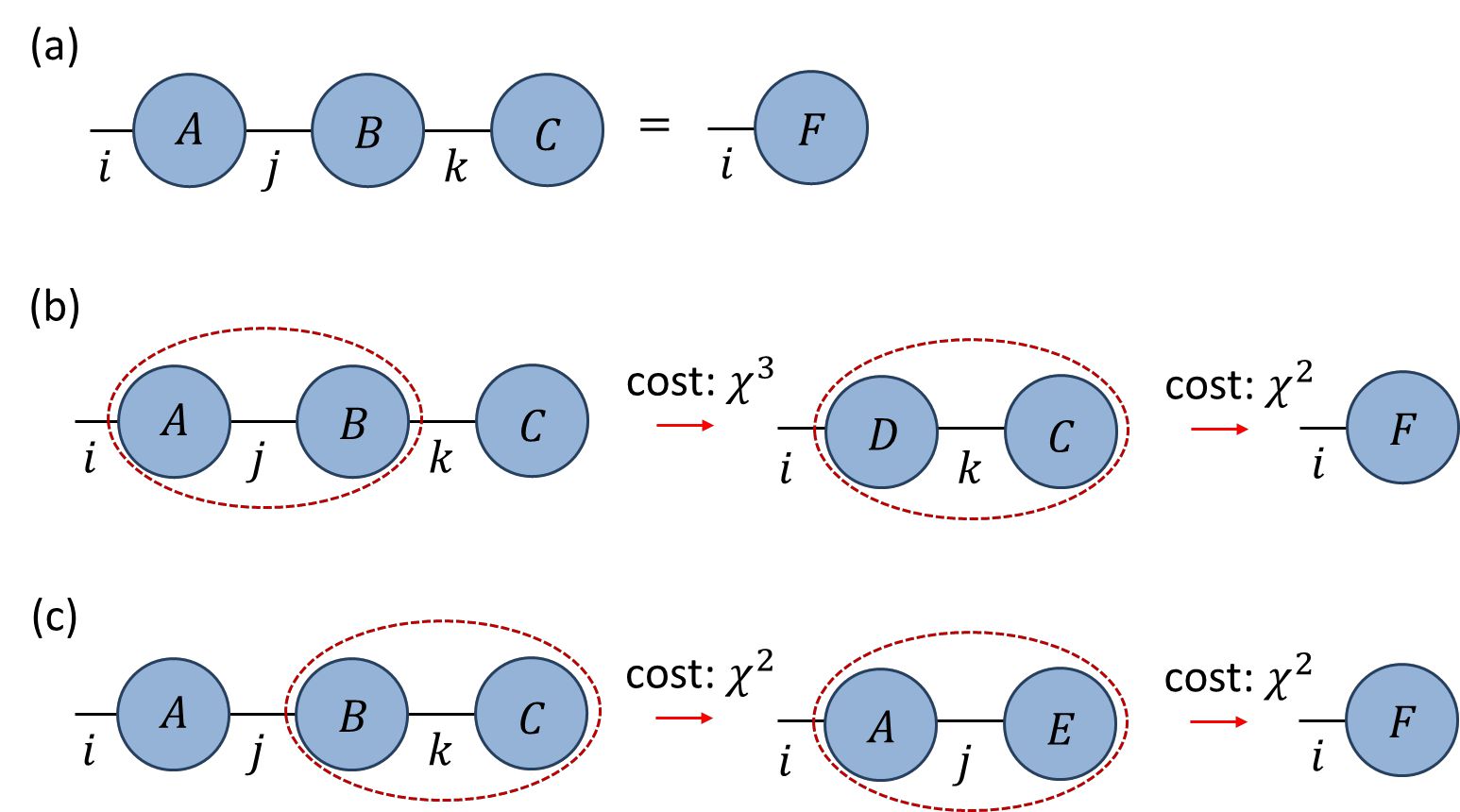}
\caption{(a) A product of three tensors $\{A,B,C \}$ is contracted to a tensor $F$, where all indices are assumed to be of dimension $\chi$. (b-c) The total computational cost of contracting the network depends on the sequence of pairwise contractions; the cost from following the sequence in (b) scales as $(\chi^3 + \chi^2)$ as compared to the cost from (c) which scales as $(2\chi^2)$.}
\label{fig:costs}
\end{center}
\end{figure}

\subsection{Contraction sequence}
It is straight-forward to establish that, when breaking a network contraction into a sequence of binary contractions, the choice of sequence can affect the total computational cost. As an example, we consider the product of two matrices $A$, $B$ with vector $C$, 
\begin{equation}
{F_i} = \sum\limits_{j,k} {{A_{ij}}{B_{jk}}{C_k}}, \label{eq:s5}
\end{equation}
where all indices are assumed to be dimension $\chi$, see also Fig. \ref{fig:costs}. If we evaluate this expression by first performing the matrix-matrix multiplication, i.e. as $F = (A\times B)\times C$, then the leading order computational cost scales as $O(\chi^3)$ by Eq. \ref{eq:s4}. Alternatively, if we evaluate the expression by first performing the matrix-vector multiplication, i.e. as $F = A\times (B\times C)$, then the leading order computational cost scales as $O(\chi^2)$. Thus it is evident that the sequence of binary contractions needs to be properly considered in order to minimize the overall computational cost.

So how does one find the optimal contraction sequence for some tensor network of interest? For the networks that arise in common algorithms (such as DMRG, MERA, PEPS, TRG and TNR) it is relatively easy, with some practice, to find the optimal sequence through manual inspection or trial-and-error. This follows as most networks one needs to evaluate contain fewer than 10 tensors and the tensor index dimensions take a only single or a few distinct values within a network, which limits the number of viable contraction sequences that need be considered. More generally, determination of optimal contraction sequences is known to be an NP-hard problem \cite{Order1}, such that it is very unlikely that an algorithm which scales polynomially with the number of tensors in the network will ever be found to exist. However, numerical methods based on exhaustive searches and/or heuristics can typically find optimal sequences for networks with fewer than 20 tensors in a reasonable amount of time \cite{Order1, Order2, Order4, Order5}, and larger networks are seldom encountered in practice. 

Note that many tensor network optimisation algorithms based on an iterative sweep, where the same network diagrams are contracted each iteration (although perhaps containing different tensors and with different bond dimensions). The usual approach in this setting is to determine the optimal sequences once, before beginning the iterative sweeps, using the initial bond dimensions and then cache the sequences for re-use in later iterations. The contraction sequences are then only recomputed the if the bond dimensions stray too far from the initial values.

\subsection{Network contraction routines}
Although certainly feasible, manually writing the code for each tensor network contraction as a sequence of pairwise contractions is not recommended. Not only is substantial programming effort required, but this also results in code which is error-prone and difficult to check. There is also a more fundamental problem: contracting a network by manually writing a sequence of binary contractions requires specifying a particular contraction sequence at the time of coding. However, in many cases the index dimensions within networks are variable, and the optimal sequence can change depending on the relative sizes of dimensions. For instance, one may have a network which contains indices of dimensions $\chi_1$ and $\chi_2$, where the optimal contraction sequence changes dependant of whether $\chi_1$ is larger or smaller than $\chi_2$. In order to have a program which works efficiently in both regimes, one would have to write code separately for both contraction sequences.
 
Given the considerations above, the use of an automated contraction routine, such as the `\texttt{ncon}' (Network-CONtractor) function \cite{Order3,TenNet} or something similar from an existing tensor network library \cite{TB1, TB2, TB3, TB4, TB5, TB6, TB7, TB8}, is highly recommended. Automated contraction routines can evaluate any network contraction in a single call by appropriately generating and evaluating a sequence of binary contractions, hence greatly reducing both the programming effort required and the risk of programming errors occurring. Most contraction routines, such as `\texttt{ncon}', also remove the need to fix a contraction sequence at the time of writing the code, as the sequence can be specified as an input variable to the routine and thus can be changed without the need to rewrite any code. This can also allow the contraction sequence to be adjusted dynamically at run-time to ensure that the sequence is optimal given the specific index dimensions in use.

\subsection{Summary: contractions}
In evaluating a network of multiple tensors, it is always more efficient to break the contraction into a sequence of pairwise tensor contractions, each of which should (usually) be recast into a matrix-matrix multiplication in order to achieve optimal computational performance. The total cost of evaluating a network can depend on the particular sequence of pairwise contractions chosen. While there is no known method for determining an optimal contraction sequence that is efficiently scalable in the size of the network, manual inference or brute-force numeric searches are usually viable for the relatively small networks encountered in common tensor network algorithms. When coding a tensor network program it is useful to utilise an automated network contraction routine which can evaluate a tensor network in a single call by properly chaining together a sequence of pairwise contractions. This not only reduces the programming effort required, but also grants a program more flexibility in allowing a contraction sequence to be easily changed.

\begin{figure}[!t!b]
\begin{center}
\includegraphics[width=8cm]{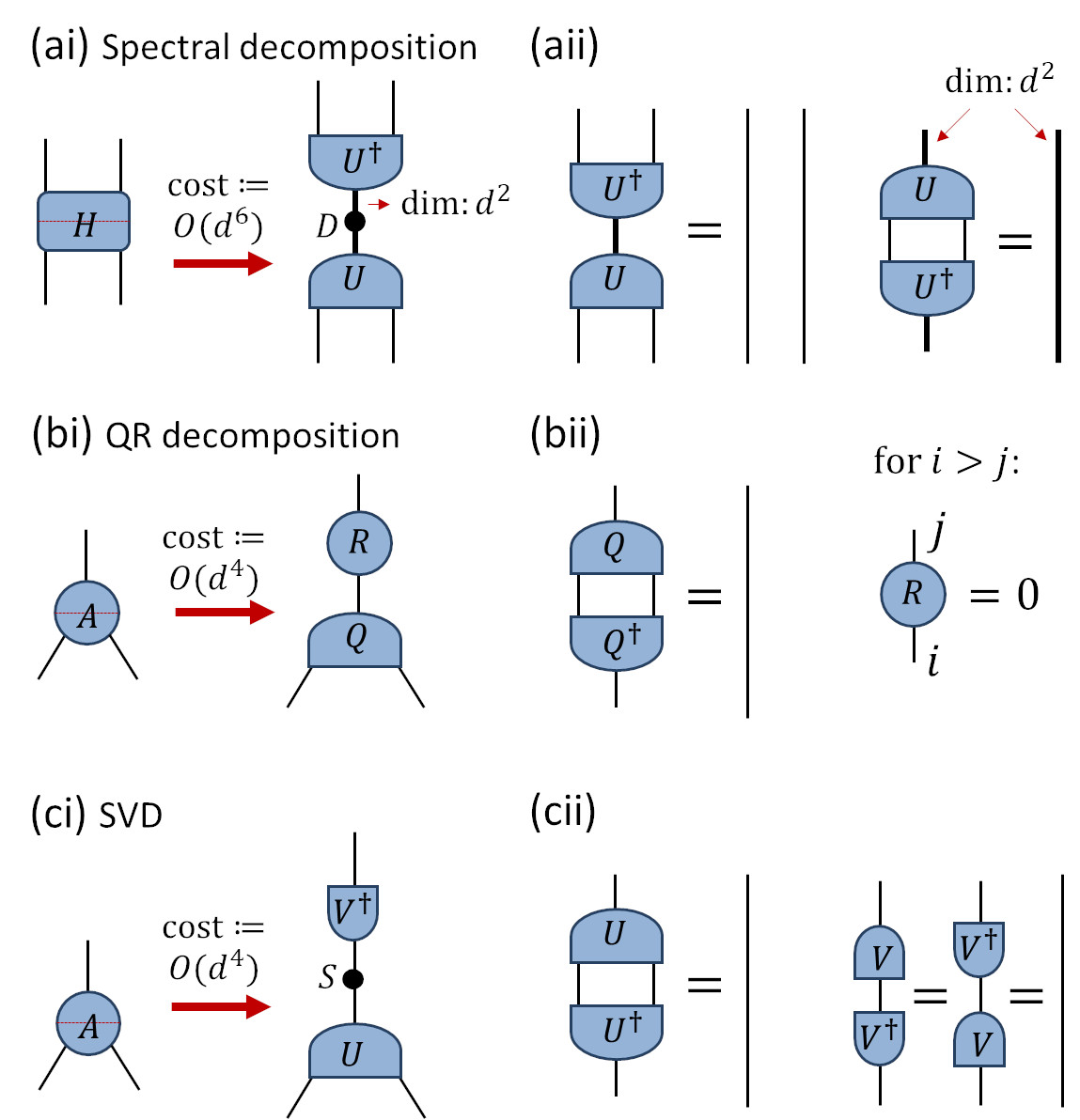}
\caption{Depiction of some common matrix decompositions. All indices are assumed to be of dimension $d$ unless otherwise indicated. (a-i) The spectral decomposition is applied to the order-4 Hermitian tensor $H$ across the partition indicated by the dashed line, yielding a diagonal matrix of eigenvalues $D$ and a unitary $U$. (a-ii) The unitary tensor $U$ annihilates to identity with its conjugate, as per Eq.\ref{eq:g1}. (b-i) The QR decomposition is applied to the order-3 tensor $A$ across the partition indicated, yielding an isometry $Q$ and an upper triangular matrix $R$. (b-ii) The isometry $Q$ annihilates to identity with its conjugate as per Eq.\ref{eq:g2}, while the $R$ matrix is upper triangular. (c-i) The singular value decomposition (SVD) is applied to the order-3 tensor $A$ across the partition indicated, yielding an isometry $U$, a diagonal matrix of singular values $S$, and a unitary $V$. (c-ii) Depiction of the constraints satisfied by the isometry $U$ and unitary $V$.}
\label{fig:S4}
\end{center}
\end{figure}

\section{Matrix Factorizations} \label{sect:Decomp}
Another key operation common in tensor network algorithms, complimentary to the tensor contractions considered previously, are factorizations. In this section we will discuss some of the various means by which a higher-order tensor can be split into a product of fewer-order tensors. In particular, the means that we consider involve applying standard matrix decompositions\cite{Mat1, Mat2}, to tensor unfoldings, such that this section may serve as a review of the linear algebra necessary before consideration of more sophisticated network decompositions. Specifically we recount the spectral decomposition, QR decomposition and singular value decomposition and outline their usefulness in the context of tensor networks, particular in achieving optimal low-rank tensor approximations. Before discussing decompositions, we define some special types of tensor.



\subsection{Special tensor types}
A $d$-by-$d$ matrix $U$ is said to be unitary if it has orthonormal rows and columns, which implies that it annihilates to the identity when multiplied with its conjugate, 
\begin{equation}
{U^\dag }U = U{U^\dag } = I, \label{eq:g1}
\end{equation}
where $I$ is the $d$-by-$d$ identity matrix. We define a tensor (whose order is greater than 2) as unitary with respect to a particular bi-partition of indices if the tensor can be \emph{reshaped} into a unitary matrix according to this partition. Similarly an $d_1$-by-$d_2$ matrix $W$, with $d_1 > d_2$ is said to be an isometry if 
\begin{equation}
{W^\dag }W = I, \label{eq:g2}
\end{equation}
with $I$ the $d_2$-by-$d_2$ identity matrix. Likewise we say that a tensor (order greater than 2) is isometric with respect to a particular bi-partition of indices if the tensor can be reshaped into a isometric matrix. Notice that, rather than equalling identity, the reverse order product does now evaluate to a projector $P$, 
\begin{equation}
W {W^\dag } = P, \label{eq:g3}
\end{equation}
where projectors are defined as Hermitian matrices that square to themselves, 
\begin{equation}
P = P^\dag,\; \; P^2 = P. \label{eq:g4}
\end{equation}

\subsection{Useful matrix decompositions:}
A commonly used matrix decomposition is the spectral decomposition (or eigen-decomposition). In the context of tensor network codes it is most often used for Hermitian, positive semi-definite matrices, such as for the density matrices used to describe quantum states. If $H$ is a $d\times d$ Hermitian matrix, or tensor that can be reshaped into such, then the spectral decomposition yields
\begin{equation}
H = U D{U^\dag }, \label{eq:g5}
\end{equation}
where $U$ is $d\times d$ unitary matrix and $D$ is diagonal matrix of eigenvalues, see also Fig.\ref{fig:S4}(a). The numerical cost of performing the decomposition scales as $O(d^3)$. In the context of tensor network algorithms the spectral decomposition is often applied to approximate a Hermitian tensor with one of smaller rank, as will be discussed in Sect. \ref{sect:optimal}.

Another useful decomposition is the QR decomposition. If $A$ be an arbitrary $d_1 \times d_2$ matrix with $d_1 > d_2$, or tensor that can be reshaped into such, then the QR decomposition gives
\begin{equation}
A = QR, \label{eq:g6}
\end{equation}
see also Fig.\ref{fig:S4}(b). Here $Q$ is $d_1 \times d_2$ isometry, such that $Q^\dag Q = I$, where $I$ is the $d_2 \times d_2$ identity matrix, and $R$ is $d_2 \times d_2$ upper triangular matrix. Note that we are considering the so-called economical decomposition (which is most often used in tensor network algorithms); otherwise the full decomposition gives $Q$ as a $d_1 \times d_1$ unitary and $R$ is dimension $d_1 \times d_2$. The numerical cost of the economical QR decomposition scales as the larger dimension times the square of the smaller dimension $O(d_1 {d_2}^2)$, as opposed to cost $O({d_1}^2 d_2)$ for the full decomposition. The QR decomposition is one of the most computationally efficient ways to obtain an orthonormal basis for the column space of a matrix, thus a common application is in orthogonalizing tensors within a network (i.e. transforming them into isometries), which will be discussed further in Sect. \ref{sect:pull}. 

The final decomposition that we consider is the singular value decomposition (SVD), which is also widely used in many areas of mathematics, statistics, physics and engineering. The SVD allows an arbitrary $d_1 \times d_2$ matrix $A$, where we assume for simplicity that $d_1 \ge d_2$, to be decomposed as
\begin{equation}
A = U S V^\dag \label{eq:g7}
\end{equation}
where $U$ is $d_1 \times d_2$ isometry (or unitary if $d_1 = d_2$), $S$ is diagonal $d_2 \times d_2$ matrix of positive elements (called singular values), and $V$ is $d_2 \times d_2$ unitary matrix, see also Fig.\ref{fig:S4}(c). Similar to the economical QR decomposition, we have also considered the economical form of the SVD; the full SVD would otherwise produce $U$ as a $d_1 \times d_1$ unitary and $S$ as a rectangular $d_1 \times d_2$ matrix padded with zeros. The numerical cost of the economical SVD scales as $O(d_1 {d_2}^2)$, identical to that of the economical QR decomposition. The rank of a tensor (across a specified bi-partition) is defined as the number of non-zero singular values that appear in the SVD. A common application of the SVD is in finding an approximation to a tensor by another of smaller rank, which will be discussed further in Sect. \ref{sect:optimal}. 

Notice that for any matrix $A$ the spectral decompositions of $AA^\dag$ and $A^\dag A$ are related to the SVD of $A$; more precisely, the eigenvectors of $AA^\dag$ and $A^\dag A$ are equivalent to the singular vectors in $U$ and $V$ respectively of the SVD in Eq. \ref{eq:g7}. Furthermore the (non-zero) eigenvalues in $AA^\dag$ or $A^\dag A$ are the squares of the singular values in $S$. It can also be seen that, for a Hermitian positive semi-definite matrix $H$, the spectral decomposition is equivalent to an SVD.

\begin{figure}[!t!b]
\begin{center}
\includegraphics[width=6cm]{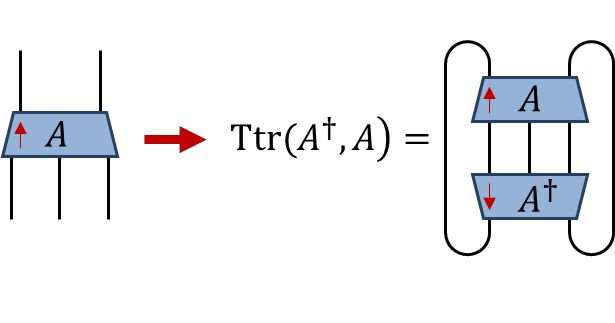}
\caption{For any tensor $A$ the tensor trace $\textrm{Ttr}$ of $A$ with its conjugate $A^\dag$ (drawn with opposite vertical orientation) is obtained by contracting over all matching indices. The Frobenius norm can be defined as the root of this tensor trace, see Eq. \ref{eq:g9}.}
\label{fig:tnorm}
\end{center}
\end{figure}

\subsection{Tensor Norms}
The primary use for matrix decompositions, such as the SVD, in the context of tensor networks is in accurately approximating a higher-order tensor as a product lower-order tensors. However, before discussing tensor approximations, it is necessary to define the tensor norm in use. A tensor norm that is particularly convenient is the Frobenius norm (or Hilbert-Schmidt norm). Given a tensor $A_{i j k\ldots}$ the Frobenius norm for $A$, denoted as $\left\| A \right\|$, is defined as the square-root of the sum of the magnitude of each element squared,
\begin{equation}
\left\| A \right\| = \sqrt {\sum\limits_{i j k\ldots} {{{\left| {{A_{i j k\ldots}}} \right|}^2}} }.  \label{eq:g8}
\end{equation}
This can be equivalently expressed as the tensor trace of $A$ multiplied by its conjugate,
\begin{equation}
\left\| A \right\| = \sqrt {\textrm{Ttr}\left( {{A^\dag } , A} \right)}, \label{eq:g9}
\end{equation}
where the tensor trace, $\textrm{Ttr}(A^\dag , A)$, represents the contraction of tensor $A$ with its conjugate over all matching indices, see Fig. \ref{fig:tnorm}. It also follows that Frobenius norm is related to the singular values $s_k$ of $A$ across any chosen bi-partition,
\begin{equation}
\left\| A \right\| = \sqrt {\sum\limits_k {{{({s_k})}^2}} }.
\end{equation}
Notice that Eq. \ref{eq:g9} implies that the difference ${\varepsilon} = \left\| {A - B} \right\| $ between two tensors $A$ and $B$ of equal dimension can equivalently be expressed as
\begin{equation}
{\left\| {A - B} \right\|^2} = \textrm{Ttr}({A^\dag },A) - 2\left| {\textrm{Ttr}({A^\dag },B)} \right| + \textrm{Ttr}({B^\dag },B). \label{eq:g10}
\end{equation}

\begin{figure}[!t!b]
\begin{center}
\includegraphics[width=9cm]{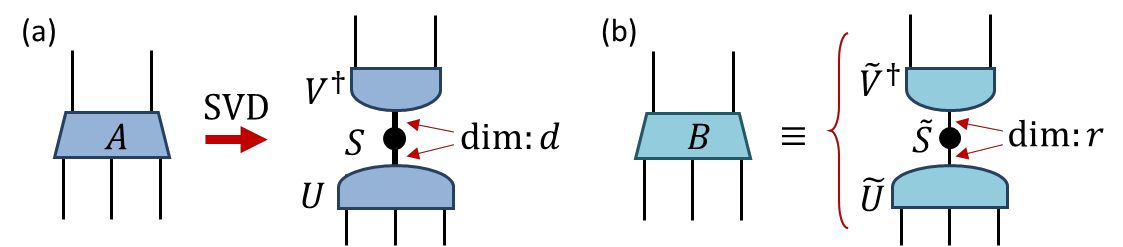}
\caption{(a) The singular value decomposition is taken on tensor $A$ across a bi-partition between its top two and bottom three indices, and is assumed to produce $d$ non-zero singular values. (b) Tensor $B$ is now defined by truncating the matrix of singular values $S \rightarrow \tilde S$ to retain only $r<d$ of the largest singular values, while similarly truncating the matrices of singular vectors, $U \rightarrow \tilde U$ and $V \rightarrow \tilde V$, to retain only the corresponding singular vectors. By the Eckart–Young–Mirsky theorem \cite{Low} it is known that $B$ is the optimal rank-$r$ approximation to $A$ (across the chosen bi-partition of tensor indices).}
\label{fig:redrank}
\end{center}
\end{figure}

\subsection{Optimal Low-rank Approximations} \label{sect:optimal}
Given some matrix $A$, or higher-order tensor that viewed as a matrix across a chosen bi-partition of its indices, we now focus on the problem of finding the tensor $B$ that best approximates $A$ according to the Frobenius norm (i.e. that which minimizes the difference in Eq. \ref{eq:g10}), assuming $B$ has a fixed rank $r$. Let us assume, without loss of generality, that tensor $A$ is equivalent to a $d_1 \times d_2$ matrix (with $d_1 \ge d_2$) under a specified bi-partition of its indices, and that $A$ has singular value decomposition,
\begin{equation}
{A_{i j}} = \sum\limits_{k=1}^{d_2} {{U_{i k}}{s_k}V_{k j}^*}, \label{eq:lowrank}
\end{equation}
where the singular values are assumed to be in descending order, $s_k \ge s_{k+1}$. Then the optimal rank $r$ tensor $B$ that approximates $A$ is known from the Eckart–Young–Mirsky theorem \cite{Low}, which states that $B$ is given by truncating to retain only the $r$ largest singular values and their corresponding singular vectors,
\begin{equation}
B_{i j} = \sum\limits_{k = 1}^{r} U_{i k} s_k V_{k j}^*.
\end{equation}
It follows that the error of approximation $\varepsilon = \left\| A - B \right\|$, as measured in the Frobenius norm, is related to the discarded singular values as
\begin{equation}
\varepsilon  = \sqrt {\sum\limits_{k > r } {{{({s_k})}^2}} }.
\end{equation}
If the spectrum of singular values is sharply decaying then the error is well approximated by the largest of the discarded singular values, $\varepsilon \approx s_{(r+1)}$.

Notice that, in the case that the tensor $A$ under consideration is Hermitian and positive definite across the chosen bi-partition, that the spectral decomposition could instead be used in Eq. \ref{eq:lowrank}. The low-rank approximation obtained by truncating the smallest eigenvalues would still be guaranteed optimal, as the spectral decomposition is equivalent to the SVD in this case.

\subsection{Summary: decompositions}
In this section we have described how special types of matrices, such as unitary matrices and projections, can be generalized to the case of tensors (which can always be viewed as a matrix across a chosen bi-partition of their indices). Similarly we have shown how several concepts from matrices, such as the trace and the norm, are also generalized to tensors. Finally, we have described how the optimal low-rank approximation to a tensor can be obtained via the SVD.

\section{Gauge Freedom} \label{sect:Gauge}
All tensor networks possess gauge degrees of freedom; exploiting the gauge freedom allows one to change the tensors within a network whilst leaving the final product of the network unchanged. In this section we describe methods for manipulating the gauge degrees of freedom and discuss the utility of fixing the gauge in a specific manner. 


\subsection{Tree tensor networks}
\begin{figure}[!t!b]
\begin{center}
\includegraphics[width=7cm]{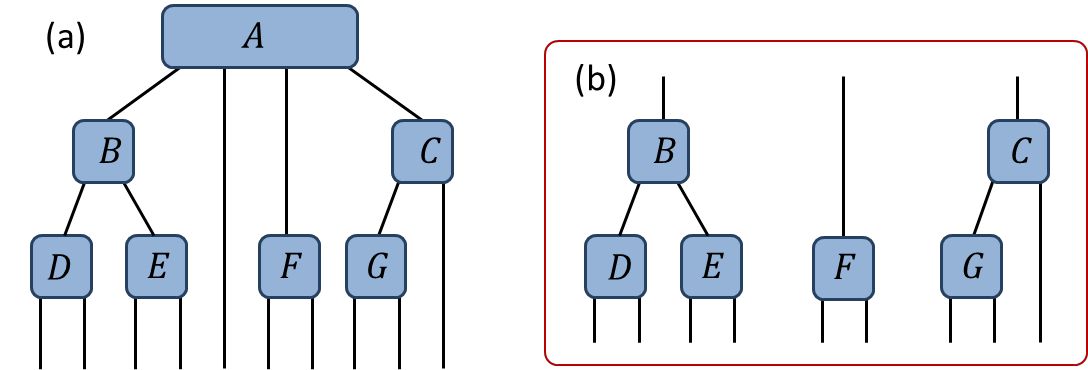}
\caption{(a) An example of a tree tensor network (TTN), here composed of 7 tensors. (b) The three (non-trivial) branches of the tree with respect to choosing $A$ as the root tensor. }
\label{fig:ttn}
\end{center}
\end{figure}

In this manuscript we shall restrict our considerations of gauge manipulations to focus only on tensors networks that do not possess closed loops (i.e. networks that correspond to acyclic graphs), which are commonly referred to as tree tensor networks (TTN) \cite{TTN1, TTN2}. Fig. \ref{fig:ttn}(a) presents an example of a tree tensor network. If we select a single tensor to act as the center (or root node) then we can understand the tree tensor network as being composed of a set of distinct branches extending from this chosen tensor. For instance, Fig. \ref{fig:ttn}(b) depicts the three branches (excluding the single trivial branch) extending from the $4^\textrm{th}$-order tensor $A$. It is important to note that connections between the different branches are not possible in networks without closed loops; this restriction is required for the re-gauging methods considered in this manuscript. However these methods can (mostly) be generalized to the case of networks containing closed loops by using a more sophisticated formalism as shown in Ref. \onlinecite{Gauge}. 

\subsection{Gauge transformations}
Consider a tensor network of multiple tensors that, under contraction of all internal indices, evaluates to some product tensor $H$. We now pose the following question: is there a different choice of tensors with the same network geometry that also evaluates to $H$? Clearly the answer to this question is yes! As shown below in Fig. \ref{fig:gaugechange}(a), on any of the internal indices of the network one can introduce a resolution of the identity (i.e. a pair of matrices $X$ and $X^{-1}$) which, by construction, does not change the final product that the network evaluates to. However, absorbing one of these matrices into each adjoining tensor changes the individual tensors while leaving the product of the network unchanged. It follows that there are infinitely many choices of tensors such that the network product evaluates to some fixed output tensor, since the gauge change matrix $X$ can be any invertible matrix. This ability to introduce an arbitrary resolution of the identity on an internal index, while leaving the product of the network unchanged, is referred to as the gauge freedom of the network.

While in some respects the gauge freedom could be considered bothersome, as it implies tensor decompositions are never unique, it can also be exploited to simplify many types of operations on tensor networks. Indeed, most tensor network algorithms require fixing the gauge in a prescribed manner in order to function correctly. In the following sections we discuss ways to fix the gauge degree of freedom as to create an \emph{orthogonality center} and the benefits of doing so.

\begin{figure}[!t!b]
\begin{center}
\includegraphics[width=7cm]{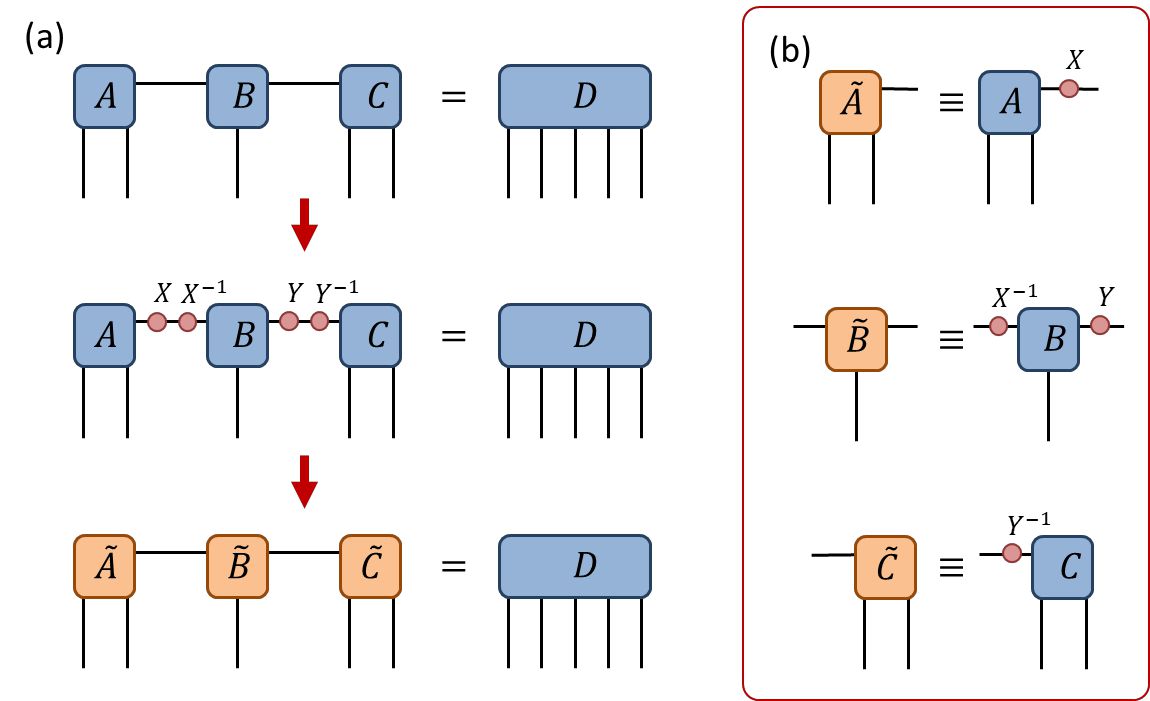}
\caption{(a) Given a network of three tensors $\{A,B,C\}$, one can introduce gauge change matrices $X$ and $Y$ (together with their inverses) on the internal indices of the network while leaving the final product $D$ of the network unchanged. (b) Definitions of the new tensors $\{\tilde A,\tilde B,\tilde C\}$ after the change of gauge.}
\label{fig:gaugechange}
\end{center}
\end{figure}

\begin{figure}[!t!b]
\begin{center}
\includegraphics[width=7cm]{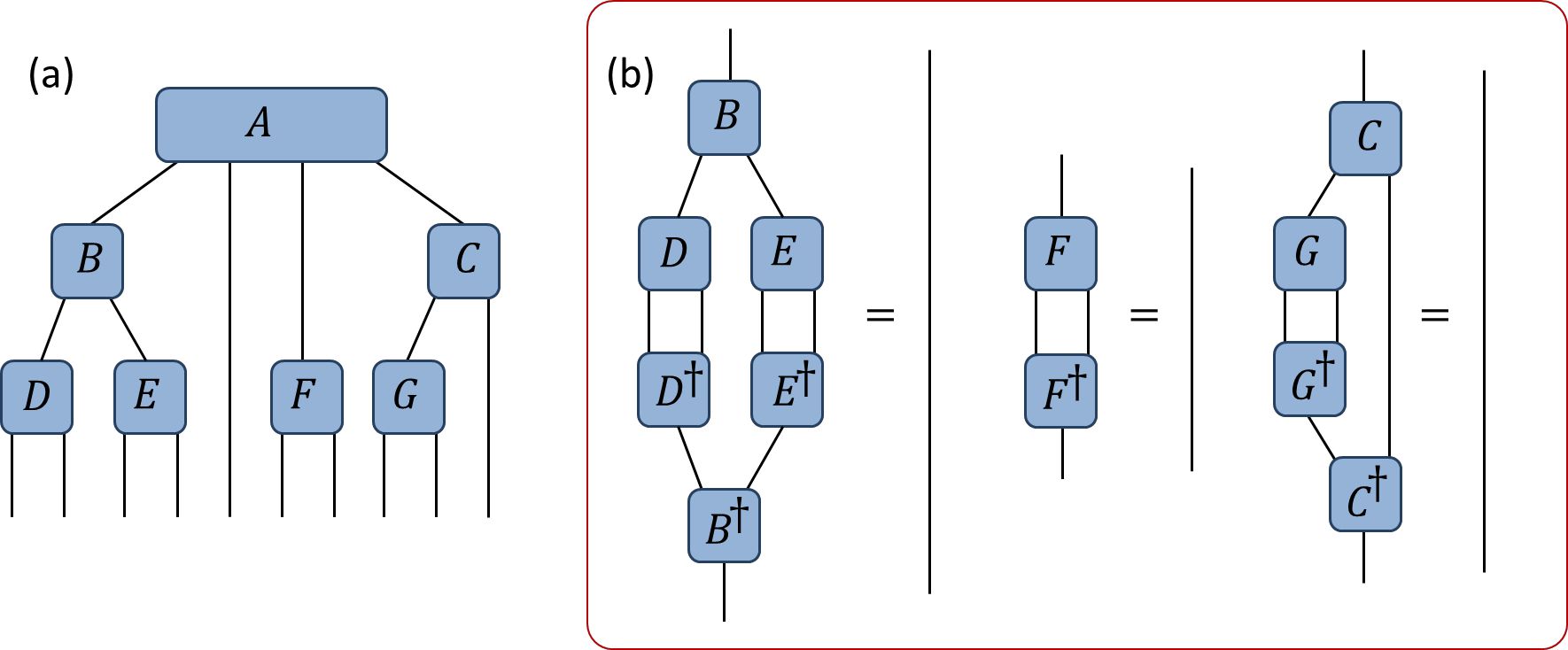}
\caption{(a) An example of a tree tensor network. (b) A depiction of the constraints required for the tensor $A$ to be an \emph{orthogonality center}: each of the branches must annihilate to the identity when contracted with its conjugate.}
\label{fig:orthogonality}
\end{center}
\end{figure}

\subsection{Orthogonality centers}
A given tensor $A$ within a network is said to be an \emph{orthogonality center} if every branch connected to tensor $A$ annihilates to the identity when contracted with its conjugate as shown in Fig. \ref{fig:orthogonality}. Equivalently, each branch must (collectively) form an isometry across the bi-partition between its open indices and the index connected to tensor $A$. By properly manipulating the gauge degrees of freedom, it is possible to turn any tensor with a tree tensor network into an orthogonality center\cite{Orth1}. We now discuss two different methods for achieving this: a `pulling through’ approach, which was a key ingredient in the original formulation of DMRG\cite{DMRG1,DMRG2,DMRG3}, and a `direct orthogonalization’ approach, which was an important part of the TEBD algorithm\cite{TEBD1, TEBD2}.

\begin{figure}[!t!b]
\begin{center}
\includegraphics[width=9cm]{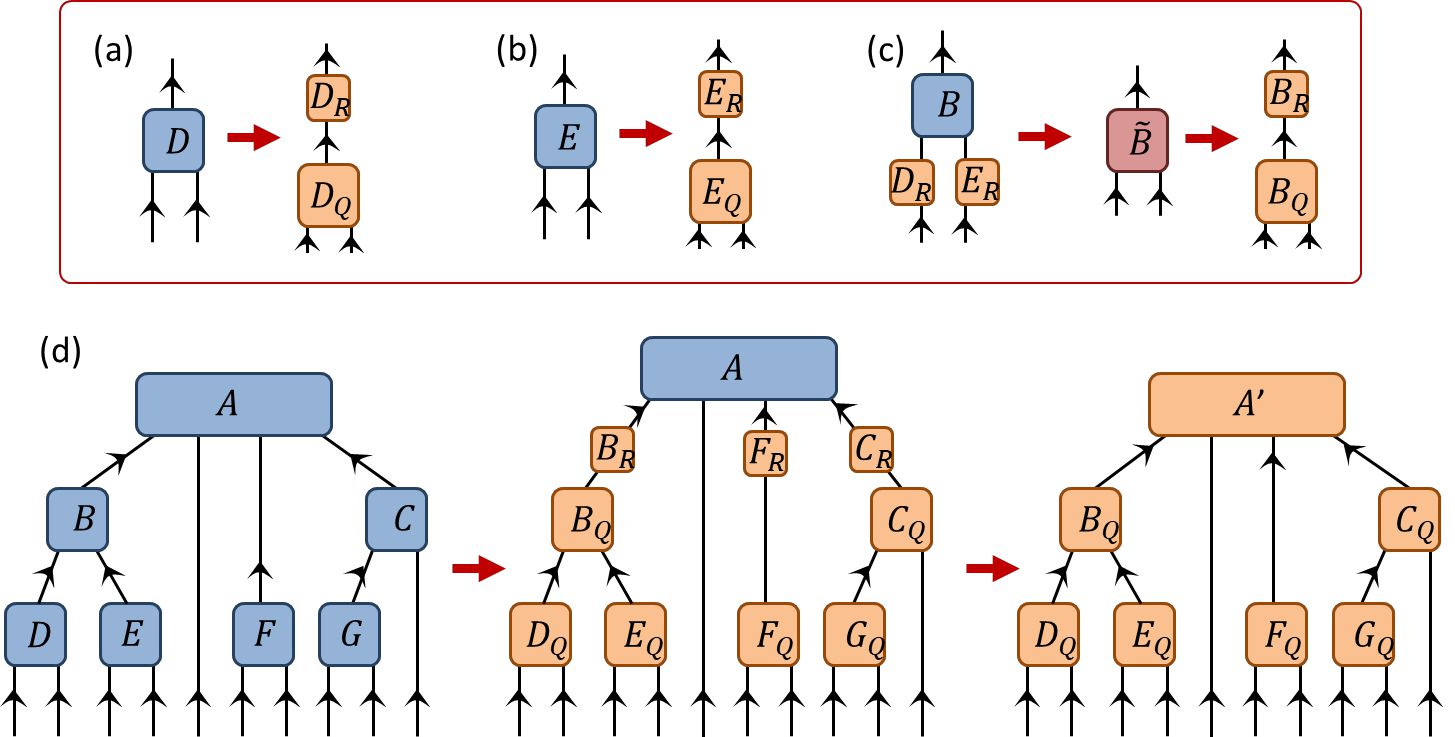}
\caption{A depiction of how the tensor $A$ can be made into an orthogonality center of the network from Fig. \ref{fig:ttn} via the ``pulling-though'' approach. (a-b) Tensors $D$ and $E$, which reside at the tips of a branch, are decomposed via the QR decomposition. (c) The $R$ components of the previous QR decompositions are absorbed into the $B$ tensor higher on the branch, which is then itself decomposed via the QR decomposition. (d) Following this procedure, all tensors in the network are orthogonalized (with respect to their incoming versus outgoing indices) such that $A'$ becomes an orthogonality center of the network. }
\label{fig:pullthrough}
\end{center}
\end{figure}

\subsubsection{Creating an orthogonality center via `pulling through’} \label{sect:pull}
Here we describe a method for setting a tensor $A$ within a network as an orthogonality center through iterative use of the QR decomposition. The idea behind his method is very simple: if each individual tensor within a branch is transformed into a (properly oriented) isometry, then the entire branch collectively becomes an isometry and thus the tensor at center of the branches becomes an orthogonality center. Let us begin by orienting each index of the network by drawing an arrow pointing towards the desired center $A$. Then, starting at the tip of each branch, we should perform a QR decomposition on each tensor based on a bi-partition between its incoming and outgoing indices. The $R$ part of the decomposition should then be absorbed into the next tensor in the branch (i.e. closer to the root tensor $A$), and the process repeated as depicted in Fig. \ref{fig:pullthrough}(a-c). At the final step an $R$ part of the QR decomposition from each branch is absorbed into the central tensor $A$ and the process is complete, see also Fig. \ref{fig:pullthrough}(d).

Note that the SVD could be used as an alternative to the QR decomposition: instead of absorbing the $R$ part of the QR decomposition into the next tensor in the branch one could absorb the product of the $S$ and $V$ part of the SVD from Eq. \ref{eq:g7}. However, in practice, the QR decomposition is most often preferable as it computationally cheaper than the SVD. 

\begin{figure}[!t!b]
\begin{center}
\includegraphics[width=8cm]{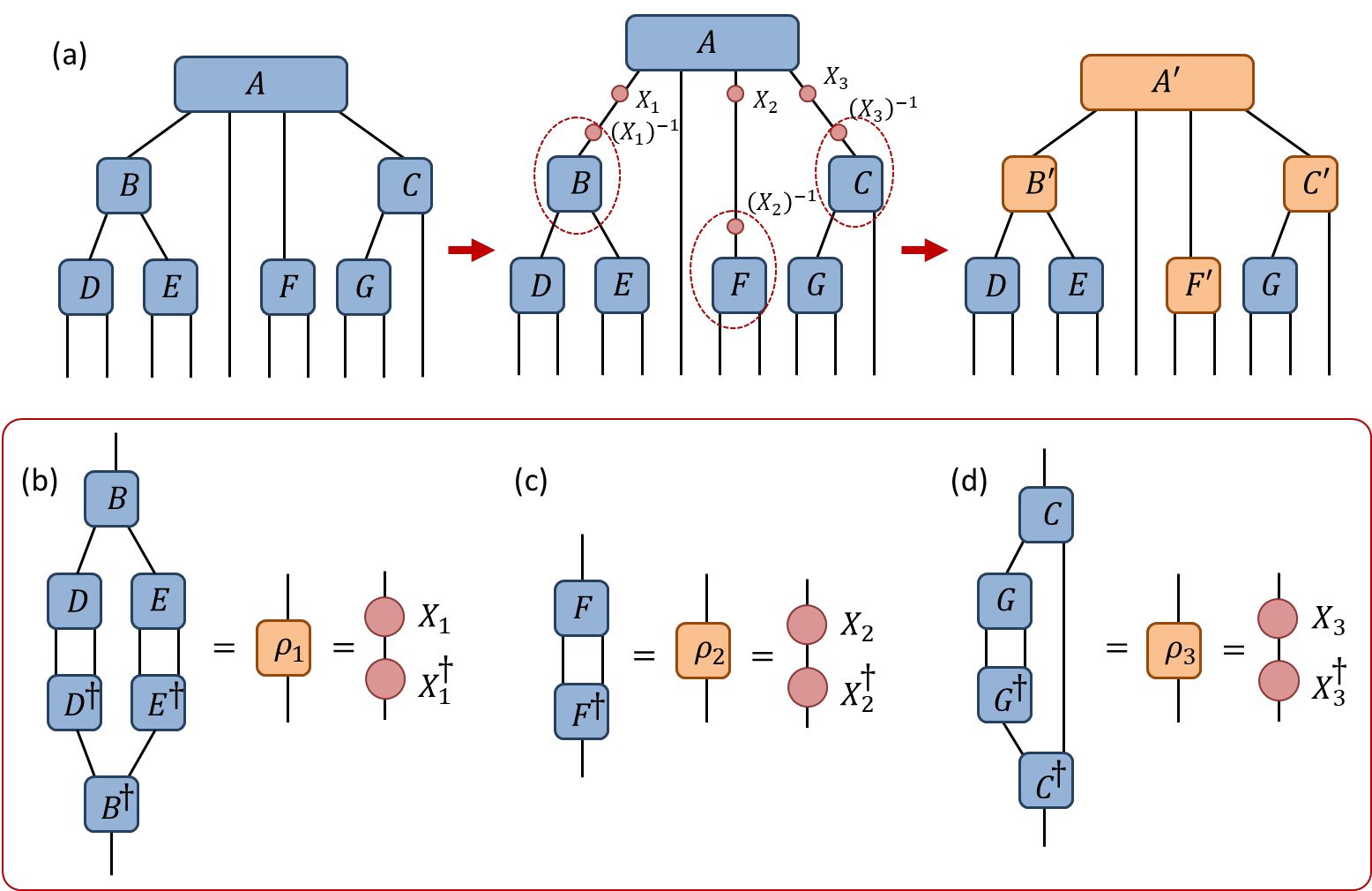}
\caption{A depiction of how the tensor $A$ from the network of Fig. \ref{fig:ttn} can be made into an orthogonality center via the ``direct orthogonalization'' approach. (a) A change of gauge is made on every (non-trivial) branch connected to $A$, such that $A$ becomes an orthogonality center. (b-d) The gauge change matrices $\{X_1, X_2, X_3 \}$ are obtained by contracting each branch with its Hermitian conjugate and then taking the principle root.}
\label{fig:direct}
\end{center}
\end{figure}

\subsubsection{Creating an orthogonality center via `direct orthogonalization’}
Here we describe a method for setting a tensor $A$ within a network as an orthogonality center based on use of a single decomposition for each branch, as depicted in Fig. \ref{fig:direct}. (i) We begin by computing the positive-definite matrix $\rho$ for each branch (with respect to the center tensor $A$) by contracting the branch with its Hermitian conjugates. (ii) The principle square root $X$ of each of the matrices $\rho$ is then computed, i.e. such that $\rho = X X^\dag$, which can be computed using the spectral decomposition if necessary. (iii) Finally, a change of gauge is made on each of the indices of tensor $A$ using the appropriate $X$ matrix and its corresponding inverse $X^{-1}$, with the $X$ part absorbed in tensor $A$ and the $X^{-1}$ absorbed in the leading tensor of the branch such that the branch matrix transforms as $\rho \rightarrow \tilde \rho = X^{-1} \rho (X^{-1})^\dag$. It follows that the tensor $A$ is now an orthogonality center as each branch matrix $\tilde \rho$ of the transformed network evaluates as the identity,
\begin{equation}
\tilde \rho = X^{-1} \rho (X^{-1})^\dag = X^{-1} X X^\dag (X^{-1})^\dag = I.
\end{equation}
Note that, for simplicity, we have assumed that the branch matrices $\rho$ do not have zero eigenvalues, such that their inverses exist. Otherwise, if zero eigenvalues are present, the current method is not valid unless the index dimensions are first reduced by truncating any zero eigenvalues.

\subsubsection{Comparison of methods for creating orthogonality centers}
Each of the two methods discussed to create an orthogonality center have their own advantages and disadvantages, such that the preferred method may depend on the specific application under consideration. In practice, the `direct orthogonalization' is typically computationally cheaper and easier to execute, since this method only requires changing the gauge on the indices connected to the center tensor. In contrast the `pulling through' method involves changing the gauge on all indices of the network. Additionally, the `direct orthogonalization' approach can easily be employed in networks of infinite extent, such as infinite MPS \cite{TEBD1, TEBD2}, if the matrix $\rho$ associated to a branch of infinite extent can be computed using by solving for a dominant eigenvector. While `pulling through' can also potentially be employed for networks of infinite extent, i.e. through successive decompositions until sufficient convergence is achieved, this is likely to be more computationally expensive. However the `pulling through' approach can be advantageous if the branch matrices $\rho$ are ill-conditioned as the errors due to floating-point arithmetic are lesser. This follows since the `direct orthogonalization' requires one to square the tensors in each branch. The `pulling-through' approach also results in transforming every tensor in the network (with the exception of the center tensor) into an isometry, which may be desirable in certain applications.

\begin{figure}[!t!b]
\begin{center}
\includegraphics[width=6cm]{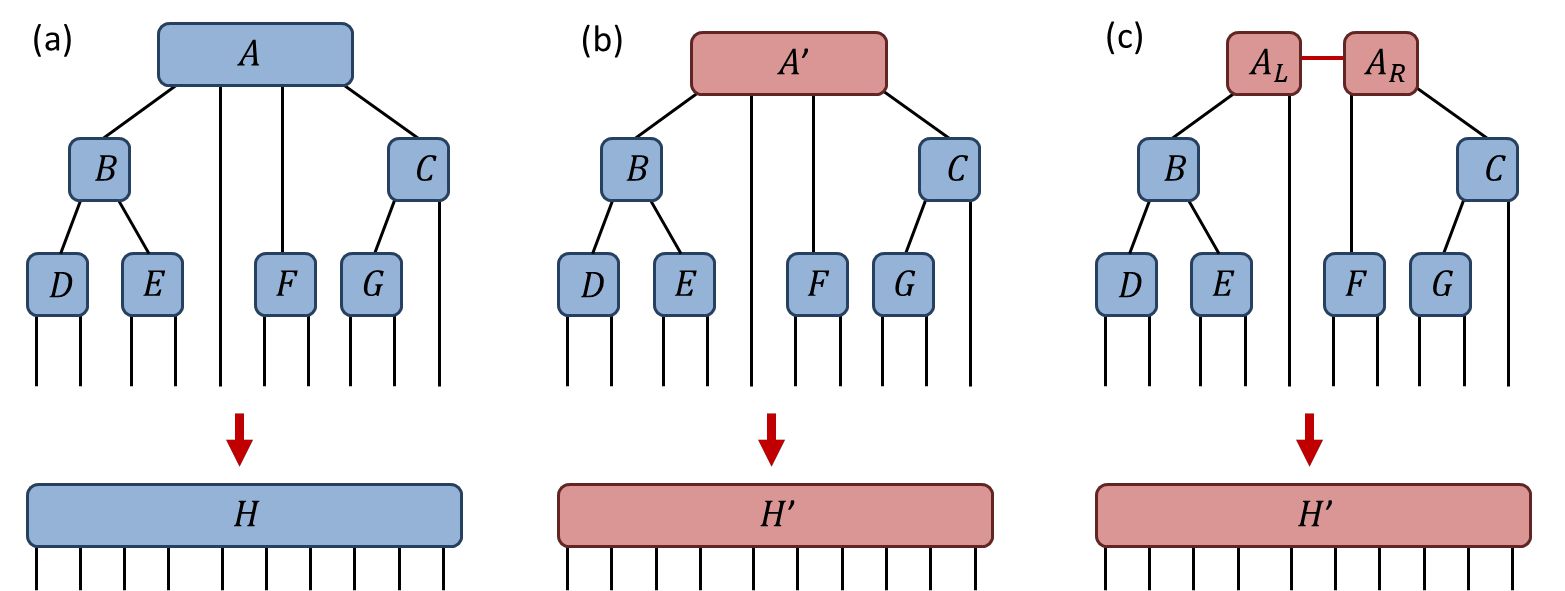}
\caption{(a) A network of 7 tensors $\{A,B,C,D,E,F,G \}$ contracts to give a tensor $H$. (b) After replacing a single tensor $A \rightarrow A'$ the network contracts to a different tensor $H'$. (c) The tensor $A'$ is decomposed into a pair of tensors $A_L$ and $A_R$, leaving the final tensor $H'$ unchanged.}
\label{fig:localglobal}
\end{center}
\end{figure}

\subsection{Decompositions of tensors within networks}
In Sect. \ref{sect:Decomp} it was described how the SVD could be applied to find the optimal low-rank approximation to a tensor in terms of minimizing the Frobenius norm between the original tensor and the approximation. In the present section we extend this concept further and detail how, by first creating an orthogonality center, a tensor within a network can be optimally decomposed as to minimize the \emph{global} error coming from consideration of the entire network. 

\begin{figure}[!t!b]
\begin{center}
\includegraphics[width=7cm]{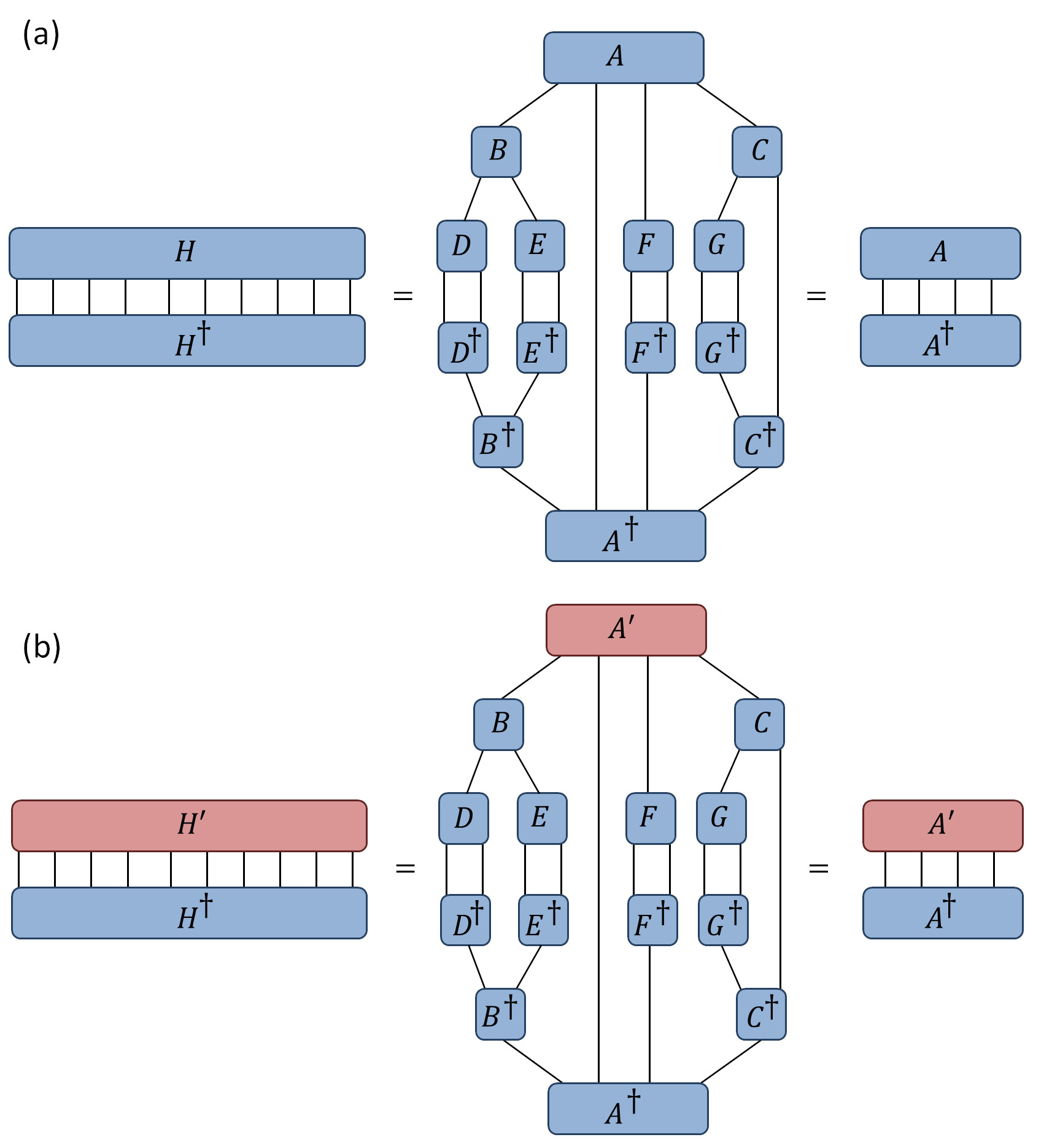}
\caption{(a) In the network from Fig. \ref{fig:localglobal}(a), if the tensor $A$ is an orthogonality center then it follows that the norm of the network $\| H \|$ is equal to the norm of the center tensor $\| A \|$. (b) Similarly it follows that, in changing only the center tensor $A\rightarrow A'$, the global overlap between the networks $H$ and $H'$ is equal to the local overlap between the center tensors $A$ and $A'$.}
\label{fig:overlap}
\end{center}
\end{figure}

Let us consider a tree tensor network of tensors $\{A,B,C,D,E,F,G \}$ that evaluates to a tensor $H$, as depicted in Fig. \ref{fig:localglobal}(a). We now replace a single tensor $A$ from this network by a new tensor $A'$ such that the network now evaluates to a tensor $H'$ as depicted in Fig. \ref{fig:localglobal}(b). Our goal is to address the following question: how can we find the optimal low-rank approximation $A'$ to tensor $A$ such that the error from the full network, $\| H - H' \|$, is minimised? Notice that if we follow the method from Sect. \ref{sect:Decomp} and simple truncate the smallest singular values of $A$, then this will only ensure that the local error, $\| A - A' \|$, is minimised. The key to resolving this issue is through creation of an orthogonality center, which can reduce the global norm of a network to the norm of a single tensor. Specifically if tensor $A$ is an orthogonality center of a network that evaluates to a final tensor $H$ then it follow from the definition of an orthogonality center that $\| H \| = \| A \| $, as depicted in Fig. \ref{fig:overlap}(a). Thus it also can be seen that under replacement of the center tensor $A$ with a new tensor $A'$, such that the network now evaluates to a new tensor $H'$, that the difference between the tensors $\| A - A' \|$ is precisely equal to the global difference between the networks $\| H - H' \|$. This follows as the overlap of $H$ and $H'$ equals the overlap of $A$ and $A'$, as depicted in Fig. \ref{fig:overlap}(b). In other words, by appropriately manipulating the gauge degrees of freedom in a network, the \emph{global} difference resulting from changing a single tensor in a network can become equivalent to the \emph{local} difference between the single tensors. The solution to the problem of finding the optimal low-rank approximation $A'$ to a tensor $A$ within a network thus becomes clear; we should first adjust the gauge such that $A$ becomes an orthogonality center, after which we can follow the method from Sect. \ref{sect:Decomp} and create the optimal global approximation (i.e. that which minimizes the global error) by truncating the smallest singular values of $A$. The importance of this result in the context tensor network algorithms cannot be overstated; this understanding for how to optimally truncate a single tensor within a tensor network, see also Ref.\onlinecite{CanForm}, is a key aspect of the DMRG algorithm \cite{DMRG1, DMRG2, DMRG3}, the TEBD algorithm \cite{TEBD1, TEBD2} and many other tensor network algorithms.   

\subsection{Summary: gauge freedom}
In the preceding section we discussed manipulations of the gauge degrees of freedom in a tensor network and described two methods that can be used to create an orthogonality center. The proper use of an orthogonality center was then demonstrated to allow one to decompose a tensor within a network in such a way as to minimize the global error. Note that while the results in this section were described only for tree tensor networks (i.e. networks based on acyclic graphs), they can be generalized to arbitrary networks by using more sophisticated methodology \cite{Gauge}. 

\section{Conclusions}
Network contractions and decompositions are the twin pillars of all tensor network algorithms. In this manuscript we have recounted the key theoretical considerations required for performing these operations efficiently and also discussed aspects of their implementation in numeric codes. We expect that a proper understanding of these results could facilitate an individuals effort to implement many common tensor network algorithms, such as DMRG, TEBD, TRG, PEPS and MERA, and also further aid researchers in the design and development of new tensor network algorithms.  

However, there are still a wide variety of additional general ideas and methods, not covered in this manuscript, that are necessary for the implementation of more advanced tensor network algorithms. These include (i) strategies for performing variational optimization, (ii) methods for dealing with decompositions in networks containing closed loops, (iii) the use of approximations in tensor network contractions. We shall address several of these topics in a follow-up work.


\begin{thebibliography}{99}


\bibitem{TN1}
J. I. Cirac and F. Verstraete, {\it Renormalization and tensor product states in spin chains and lattices}, J. Phys. A: Math. Theor. 42, 504004 (2009).

\bibitem{TN2}
G. Evenbly and G. Vidal, {\it Tensor network states and geometry}, J. Stat. Phys. {\bf 145}, 891-918 (2011).

\bibitem{TN3}
R. Orus, {\it A practical introduction to tensor networks: Matrix product states and projected entangled pair states}, Ann. Phys. 349, 117 (2014).

\bibitem{TN4}
J. C. Bridgeman and C. T. Chubb, {\it Hand-waving and Interpretive Dance: An Introductory Course on Tensor Networks}, J. Phys. A: Math. Theor. 50, 223001 (2017).

\bibitem{TN5}
S. Montangero, {\it Introduction to Tensor Network Methods - Numerical Simulations of Low-dimensional Many-body Quantum Systems}, (Springer, Berlin, 2018).

\bibitem{TN6}
R. Orus, {\it Tensor networks for complex quantum systems}, Nat. Rev. Phys. 1, 538-550 (2019).

\bibitem{TN7}
P. Silvi, F. Tschirsich, M. Gerster, J. Jünemann, D. Jaschke, M. Rizzi, and S. Montangero, {\it The Tensor Networks Anthology: Simulation techniques for many-body quantum lattice systems}, SciPost Phys. Lect. Notes 8 (2019).

\bibitem{TN8} 
S.-J. Ran, E. Tirrito, C. Peng, X. Chen, L. Tagliacozzo, G. Su, and M. Lewenstein, {\it Tensor Network Contractions Methods and Applications to Quantum Many-Body Systems}, (Springer, 2020).

\bibitem{TN9}
J. I. Cirac, D. Pérez-García, N. Schuch, and F. Verstraete, {\it Matrix product states and projected entangled pair states: Concepts, symmetries, theorems}, Rev. Mod. Phys. 93, 045003 (2021).

\bibitem{TN10}
T.G. Kolda and B.W. Bader, {\it Tensor decompositions and applications}, SIAM review, 51(3), pp.455-500 (2009).


\bibitem{Ent1} 
G. Vidal, J. I. Latorre, E. Rico, and A. Kitaev, {\it Entanglement in quantum critical phenomena}, Phys. Rev. Lett. 90, 227902 (2003).

\bibitem{Ent2}
M B Hastings, {\it An area law for one-dimensional quantum systems}, J. Stat. Mech. P08024 (2007)

\bibitem{Ent3}
J. Eisert, M. Cramer, and M. Plenio, {\it Area laws for the entanglement entropy - a review}, Rev. Mod. Phys. 82, pp. 277-306 (2010).


\bibitem{Chem1}
G. K. L. Chan and S. Sharma, {\it The density matrix renormalization group in quantum chemistry}, Annu. Rev. Phys. Chem. 62, 465 (2011). 

\bibitem{Chem2}
S. Keller,  M. Dolfi, M. Troyer, and  M. Reiher,  {\it An efficient matrix product operator representation of the quantum chemical Hamiltonian}, J. Chem. Phys. 143, 244118 (2015).

\bibitem{Chem3}
S. Szalay, M. Pfeffer, V. Murg, G. Barcza, F. Verstraete, R. Schneider, and Ö. Legeza, {\it Tensor product methods and entanglement optimization for ab initio quantum chemistry}, Int. J. Quantum Chem. 115, 1342 (2015).

\bibitem{Chem4}
G. K.-L. Chan, A. Keselman, N. Nakatani, Z. Li, and S. R. White, {\it Matrix product operators, matrix product states, and ab initio density matrix renormalization group algorithms}, J. Chem. Phys. 145, 014102 (2016).

\bibitem{Chem5}
H. Zhai and G. K.-L. Chan, {\it Low communication high performance ab initio density matrix renormalization group algorithms}, J. Chem. Phys. 154, 224116 (2021).

\bibitem{Holo1}
B. Swingle, {\it Entanglement renormalization and holography}, Phys. Rev. D 86, 065007 (2012).

\bibitem{Holo2}
M. Miyaji, T. Numasawa, N. Shiba, T. Takayanagi, and K. Watanabe, {\it Continuous Multiscale Entanglement Renormalization Ansatz as Holographic Surface-State Correspondence}, Phys. Rev. Lett. 115, 171602 (2015).

\bibitem{Holo3}
F. Pastawski, B. Yoshida, D. Harlow, and J. Preskill, {\it Holographic quantum error-correcting codes: Toy models for the bulk/boundary correspondence}, JHEP 06 (2015) 149.

\bibitem{Holo4}
P. Hayden, S. Nezami, X.-L. Qi, N. Thomas, M. Walter, and Z. Yang, {\it Holographic duality from random tensor networks}, JHEP 11 (2016) 009.

\bibitem{Holo5}
B. Czech, L. Lamprou, S. McCandlish, and J. Sully, {\it Tensor networks from kinematic space}, JHEP 07 (2016) 100.

\bibitem{Holo6}
G. Evenbly, {\it Hyperinvariant Tensor Networks and Holography}, Phys. Rev. Lett. 119, 141602 (2017).

\bibitem{ML1}
E. M. Stoudenmire and D. J. Schwab, {\it Supervised learning with tensor networks}, Adv. Neural Inf. Process. Syst. 29, pp. 4799–4807 (2016).

\bibitem{ML2}
J. Martyn, G. Vidal, C. Roberts, and S. Leichenauer, {\it Entanglement and tensor networks for supervised image classification}, arXiv:2007.06082 (2020).

\bibitem{ML3}
S. Cheng, L. Wang, and P. Zhang, {\it Supervised learning with projected entangled pair states}, Phys. Rev. B 103, 125117 (2021).

\bibitem{ML4}
J. Liu, S. Li, J. Zhang, and P. Zhang, {\it Tensor networks for unsupervised machine learning}, arXiv:2106.12974 (2021).

\bibitem{ML5}
Y. Liu, W-J Li, X. Zhang, M. Lewenstein, G. Su, and S.J Ran, {\it Entanglement-Based Feature Extraction by Tensor Network Machine Learning}, Front. Appl. Math. Stat. 7 (2021). 

\bibitem{QC1}
E. S. Fried, N. P. D. Sawaya, Y. Cao, I. D. Kivlichan, J. Romero, and A. Aspuru-Guzik, {\it qTorch: The quantum tensor contraction handler}, PLoS ONE 13(12): e0208510. (2018).

\bibitem{QC2}
B. Villalonga, S. Boixo, B. Nelson, C. Henze, E. Rieffel, R. Biswas and S. Mandrà, {\it A flexible high-performance simulator for verifying and benchmarking quantum circuits implemented on real hardware}, npj Quantum Inf 5, 86 (2019).

\bibitem{QC3}
R. Schutski, D. Lykov, and I. Oseledets, {\it Adaptive algorithm for quantum circuit simulation}, Phys. Rev. A 101, 042335 (2020).

\bibitem{QC4}
F. Pan and P. Zhang, {\it Simulation of Quantum Circuits Using the Big-Batch Tensor Network Method}, Phys. Rev. Lett. 128, 030501 (2022)

\bibitem{QC5}
M. Levental, {\it Tensor Networks for Simulating Quantum Circuits on FPGAs}, arXiv:2108.06831 (2021).

\bibitem{QC6}
T. Vincent, L. J. O'Riordan, M. Andrenkov, J. Brown, N. Killoran, H. Qi, and I. Dhand, {\it Jet: Fast quantum circuit simulations with parallel task-based tensor-network contraction}, arXiv:2107.09793 (2021).


\bibitem{Rev1} 
G. Evenbly and G. Vidal, {\it Algorithms for entanglement renormalization}, Phys. Rev. B 79, 144108 (2009).

\bibitem{Rev2} 
H.-H. Zhao, Z.-Y. Xie, Q.-N. Chen, Z.-C. Wei, J. W. Cai, and T. Xiang, {\it Renormalization of tensor-network states}, Phys. Rev. B 81, 174411 (2010).

\bibitem{Rev3}
U. Schollwoeck, {\it The density-matrix renormalization group in the age of matrix product states}, Ann. Phys. 326, 96 (2011).

\bibitem{Rev4}
H. N. Phien, J. A. Bengua, H. D. Tuan, P. Corboz, and R. Orus, {\it The iPEPS algorithm, improved: fast full update and gauge fixing}, Phys. Rev. B 92, 035142 (2015).

\bibitem{Rev5} 
G. Evenbly, {\it Algorithms for tensor network renormalization}, Phys. Rev. B 95, 045117 (2017).


\bibitem{TB1}
M. Fishman, E. M. Stoudenmire, and S. R. White, {\it ITensor Library}, \url{http://itensor.org}.

\bibitem{TB2}
Y.-J. Kao, Y.-D. Hsieh, and P. Chen, {\it Uni10: an open-source library for tensor network algorithms}, \url{https://uni10.gitlab.io/}.

\bibitem{TB3}
J. Haegeman, {\it TensorOperations}, \url{https://github.com/Jutho/TensorOperations.jl}.

\bibitem{TB4}
J. Hauschild and F. Pollmann, {\it Efficient numerical simulations with Tensor Networks: Tensor Network Python (TeNPy)}, SciPost Phys. Lect. Notes 5 (2018).

\bibitem{TB5}
S. Al-Assam, S. R. Clark, and D. Jaksch, {\it The Tensor Network Theory Library}, J. Stat. Mech. 093102 (2017).

\bibitem{TB6}
R. Olivares-Amaya, W. Hu, N. Nakatani, S. Sharma, J. Yang, and G. K.-L. Chan, {\it The ab-initio density matrix renormalization group in practice}, J. Chem. Phys. 142, 034102 (2015). Accompanying software: {\it BLOCK: Density Matrix Renormalization Group Algorithms for Quantum Chemistry}, \url{https://sanshar.github.io/Block/}

\bibitem{TB7}
C. Roberts, A. Milsted, M. Ganahl, A. Zalcman, B. Fontaine, Y. Zou, J. Hidary, G. Vidal, and S. Leichenauer, {\it TensorNetwork: A Library for Physics and Machine Learning}, arXiv:1905.01330 (2019).

\bibitem{TB8}
I. V. Oseledets, {\it TT Toolbox}, \url{https://github.com/oseledets/TT-Toolbox}


\bibitem{DMRG1}
S. R. White, {\it Density matrix formulation for quantum renormalization groups}, Phys. Rev. Lett. 69, 2863 (1992).

\bibitem{DMRG2}
S. R. White, {\it Density-matrix algorithms for quantum renormalization groups}, Phys. Rev. B 48, 10345 (1993). 

\bibitem{DMRG3}
U. Schollwoeck, {\it The density-matrix renormalization group}, Rev. Mod. Phys. 77, 259 (2005).

\bibitem{TEBD1}
G. Vidal, {\it Efficient Classical Simulation of Slightly Entangled Quantum Computations}, Phys. Rev. Lett. 91, 147902 (2003). 
\bibitem{TEBD2}
G. Vidal, {\it Efficient Simulation of One-Dimensional Quantum Many-Body Systems}, Phys. Rev. Lett. 93, 040502 (2004).


\bibitem{PEPS1} 
F. Verstraete and J. I. Cirac, {\it Renormalization algorithms for quantum-many-body systems in two and higher dimensions}, arXiv:cond-mat/0407066.

\bibitem{PEPS2} 
F. Verstraete, J.I. Cirac, and V. Murg, {\it Matrix product states, projected entangled pair states, and variational renormalization group methods for quantum spin systems}, Adv. Phys. 57, 143 (2008).

\bibitem{PEPS3}
J. Jordan, R. Orus, G. Vidal, F. Verstraete, and J. I. Cirac, {\it Classical simulation of infinite-size quantum lattice systems in two spatial dimensions}, Phys. Rev. Lett. 101, 250602 (2008).


\bibitem{MERA1}
G. Vidal, {\it A class of quantum many-body states that can be efficiently simulated}, Phys. Rev. Lett. 101, 110501 (2008).


\bibitem{TRG1} 
M. Levin and C. P. Nave, {\it Tensor renormalization group approach to two-dimensional classical lattice models}, Phys. Rev. Lett. 99, 120601 (2007).

\bibitem{TRG2} 
Z.-Y. Xie, J. Chen, M. P. Qin, J. W.  Zhu, L. P. Yang, and T. Xiang, {\it Coarse-graining renormalization by higher-order singular value decomposition}, Phys. Rev. B 86, 045139 (2012).


\bibitem{TNR1} 
G. Evenbly and G. Vidal, {\it Tensor network renormalization}, Phys. Rev. Lett. 115, 180405 (2015).


\bibitem{TenNet}
G. Evenbly, {\it Tensors.net website}, \url{https://www.tensors.net}.


\bibitem{Sym1}
S. Singh, R. N. C. Pfeifer, and G. Vidal, {\it Tensor network decompositions in the presence of a global symmetry}, Phys. Rev. A 82, 050301 (2010).

\bibitem{Sym2}
S. Singh, R. N. C. Pfeifer, and G. Vidal, {\it Tensor network states and algorithms in the presence of a global U(1) symmetry}, Phys. Rev. B 83, 115125 (2011)

\bibitem{Sym3}
A. Weichselbaum, {\it Non-abelian symmetries in tensor networks: a quantum symmetry space approach}, Ann. Phys. 327, 2972-3047 (2012).

\bibitem{Sym4}
S. Sharma, {\it A general non-Abelian density matrix renormalization group algorithm with application to the $C_2$ dimer}, J. Chem. Phys. 142, 024107 (2015).

\bibitem{Sym5}
S. Keller and  M. Reiher, {\it Spin-adapted matrix product states and operators}, J. Chem. Phys. 144, 134101 (2016).

\bibitem{Sym6}
P. Nataf and F. Mila, {\it Density matrix renormalization group simulations of SU(N) Heisenberg chains using standard Young tableaus: Fundamental representation and comparison with a finite-size Bethe ansatz}, Phys. Rev. B 97, 134420 (2018).

\bibitem{Sym7}
P. Schmoll, S. Singh, M. Rizzi, and R. Orús, {\it A programming guide for tensor networks with global SU(2) symmetry}
Ann. Phys. 419, 168232 (2020).


\bibitem{Order1}
R. N. C. Pfeifer, J. Haegeman, and F. Verstraete, {\it Faster identification of optimal contraction sequences for tensor networks}, Phys. Rev. E 90, 033315 (2014).

\bibitem{Order2}
R. N. C. Pfeifer and G. Evenbly, {\it Improving the efficiency of variational tensor network algorithms}, Phys. Rev. B 89, 245118 (2014).

\bibitem{Order4}
J. M. Dudek, L. Dueñas-Osorio, and M. Y. Vardi, {\it Efficient Contraction of Large Tensor Networks for Weighted Model Counting through Graph Decompositions}, arXiv:1908.04381v2 (2019).

\bibitem{Order5}
J. Gray and S. Kourtis, {\it Hyper-optimized tensor network contraction}, Quantum 5, 410 (2021).

\bibitem{Order3}
R. N. C. Pfeifer, G. Evenbly, S. Singh, and G. Vidal, {\it NCON: A tensor network contractor for MATLAB}, arXiv:1402.0939 (2014).


\bibitem{Mat1}
R. A. Horn, C. R. Johnson, {\it Matrix analysis}, (Cambridge University Press, 1985).

\bibitem{Mat2}
R. A. Horn, C. R. Johnson, {\it Topics in matrix analysis}, (Cambridge University Press 1991).

\bibitem{Low}
C. Eckart and G. Young, {\it The approximation of one matrix by another of lower rank}, Psychometrika 1, 211–218 (1936).


\bibitem{TTN1}
Y. Shi, L. Duan and G. Vidal, {\it Classical simulation of quantum many-body systems with a tree tensor network}, Phys. Rev. A 74, 022320 (2006).

\bibitem{TTN2}
L. Tagliacozzo, G. Evenbly and G. Vidal, {\it Simulation of two-dimensional quantum systems using a tree tensor network that exploits the entropic area law}, Phys. Rev. B 80, 235127 (2009).

\bibitem{Gauge}
G. Evenbly, {\it Gauge fixing, canonical forms and optimal truncations in tensor networks with closed loops}, Phys. Rev. B 98, 085155 (2018).

\bibitem{Orth1}
S. Holtz, T. Rohwedder, and R. Schneider, {\it The Alternating Linear Scheme for Tensor Optimization in the Tensor Train Format}, SIAM J. Sci. Comput., 34(2), A683–A713. 

\bibitem{CanForm}
Y. Zhang and E. Solomonik, {\it On Stability of Tensor Networks and Canonical Forms}, arXiv:2001.01191 (2020).







\end{thebibliography}
\end{document}